\documentclass[aps,prl,twocolumn,superscriptaddress]{revtex4-1}
\usepackage{amsmath}
\usepackage{amssymb}

\usepackage{graphics}
\usepackage{graphicx}
\usepackage{float}
\usepackage{subfigure}
\usepackage{color}
\usepackage{hyperref}

\begin{document}
\title{Demonstrating the wormhole mechanism of entanglement spectrum via a perturbed boundary}
\author{Zenan Liu}
\affiliation{State Key Laboratory of Optoelectronic Materials and Technologies, Guangdong Provincial Key Laboratory of Magnetoelectric Physics and Devices, Center for Neutron Science and Technology, School of Physics, Sun Yat-Sen University, Guangzhou 510275, China}

\author{Rui-Zhen Huang}
\email{ruizhen.huang@ugent.be}
\affiliation{Department of Physics and Astronomy, Ghent University, Krijgslaan 281, S9, B-9000 Ghent, Belgium}

\author{Zheng Yan}
\email{zhengyan@westlake.edu.cn}
\affiliation{Department of Physics, School of Science, Westlake University, Hangzhou 310030, Zhejiang, China}
\affiliation{Institute of Natural Sciences, Westlake Institute for Advanced Study, Hangzhou 310024, Zhejiang, China}
\affiliation{Lanzhou Center for Theoretical Physics $\&$ Key Laboratory of Theoretical Physics of Gansu Province, Lanzhou University, Lanzhou, Gansu 730000, China}

\author{Dao-Xin Yao}
\email{yaodaox@mail.sysu.edu.cn}
\affiliation{State Key Laboratory of Optoelectronic Materials and Technologies, Guangdong Provincial Key Laboratory of Magnetoelectric Physics and Devices, Center for Neutron Science and Technology, School of Physics, Sun Yat-Sen University, Guangzhou 510275, China}
\affiliation{International Quantum Academy, Shenzhen 518048, China}

\begin{abstract}
The Li-Haldane conjecture is one of the most famous conjectures in physics and opens a new research area in the quantum entanglement and topological phase. 
Although a lot of theoretical and numerical works have confirmed the conjecture in topological states with bulk-boundary correspondence, the cases with gapped boundary and the systems in high dimension are widely unknown. What is the valid scope of the Li-Haldane conjecture? Via the newly developed quantum Monte Carlo scheme, we are now able to extract the large-scale entanglement spectrum  and study its relation with the edge energy spectrum generally. Taking the two-dimensional Affleck-Kennedy-Lieb-Tasaki  model with a tunable boundary on the square-octagon lattice as an example, we find several counter-examples which can not be explained by the Li-Haldane conjecture; $e.g.$, the low-lying entanglement spectrum does not always show similar behaviors as the energy spectrum on the virtual boundary, and sometimes the ES resembles the energy spectrum of the edge even if it is gapped.
Finally, we demonstrate that the newly proposed "wormhole mechanism" on the path integral of reduced density matrices is the formation principle of the general ES. We find that the Li-Haldane conjecture is a particular case in some limit of the wormhole picture while all the examples out of the conjecture we have studied can totally be explained within the wormhole mechanism framework. Our results provide important evidence for demonstrating that the wormhole mechanism is the fundamental principle to explain the ES.
\end{abstract}
\date{\today}
\maketitle

\section{Introduction}
In the past decades, entanglement has become a key concept in describing quantum matters, especially exotic new quantum phases from the quantum information viewpoint. It leads to a unified understanding of quantum matter and information and serves as a quintessential quantity to detect and characterize the informational, field-theoretical, and topological properties of quantum many-body states~\cite{vidal2003entanglement,Korepin2004universality,Kitaev2006,Levin2006}, which combines the conformal field theory (CFT) \cite{HOLZHEY1994,Calabrese_2004}
and the categorial description of the problem~\cite{Calabrese2008entangle,Fradkin2006entangle,Nussinov2006,Nussinov2009,CASINI2007,JiPRR2019,ji2019categorical,kong2020algebraic,XCWu2020,XCWu2021,JRZhao2021,JRZhao2022,Wang2022scaling}.  Bipartite entanglement entropy (EE) was widely used to identify quantum phases and phase transitions \cite{Roger2010,Kallin_2014,JRZhao2021,JRZhao2022}. 
For topologically ordered systems, a topological EE term was also proposed to detect the quantum dimension of the topological excitation \cite{Kitaev2006,isakov2011topological,Zhang2011EE,JRZhao2022}. 

More importantly, the entanglement spectrum (ES) contains more information than EE~\cite{Li2008entangle}. The low-lying ES has been widely employed as a fingerprint of CFT and topology in the investigation of highly entangled quantum matter~\cite{Pollmann2010entangle,Fidkowski2010,Yao2010,XLQi2012,Canovi2014,LuitzPRL2014,
LuitzPRB2014,LuitzIOP2014,Chung2014,Pichler2016,Cirac2011,Stoj2020,guo2021entanglement,
Grover2013,Assaad2014,Assaad2015,Parisen2018,yu2022conformal}. 
For topologically ordered phases with gapless boundaries, Li and Haldane pointed out that the low-lying ES of topological states will be similar to the energy spectrum of the edge state, which is dubbed the Li-Haldane conjecture and relates the low-lying ES to the energy spectrum~\cite{Li2008entangle}. Their work demonstrated that the general $\nu=5/2$ topological states have the same low-lying ES to identify the topology and CFT structure on the boundary.

\begin{figure}[htb]
 \centering
 \includegraphics[width=0.50\textwidth]{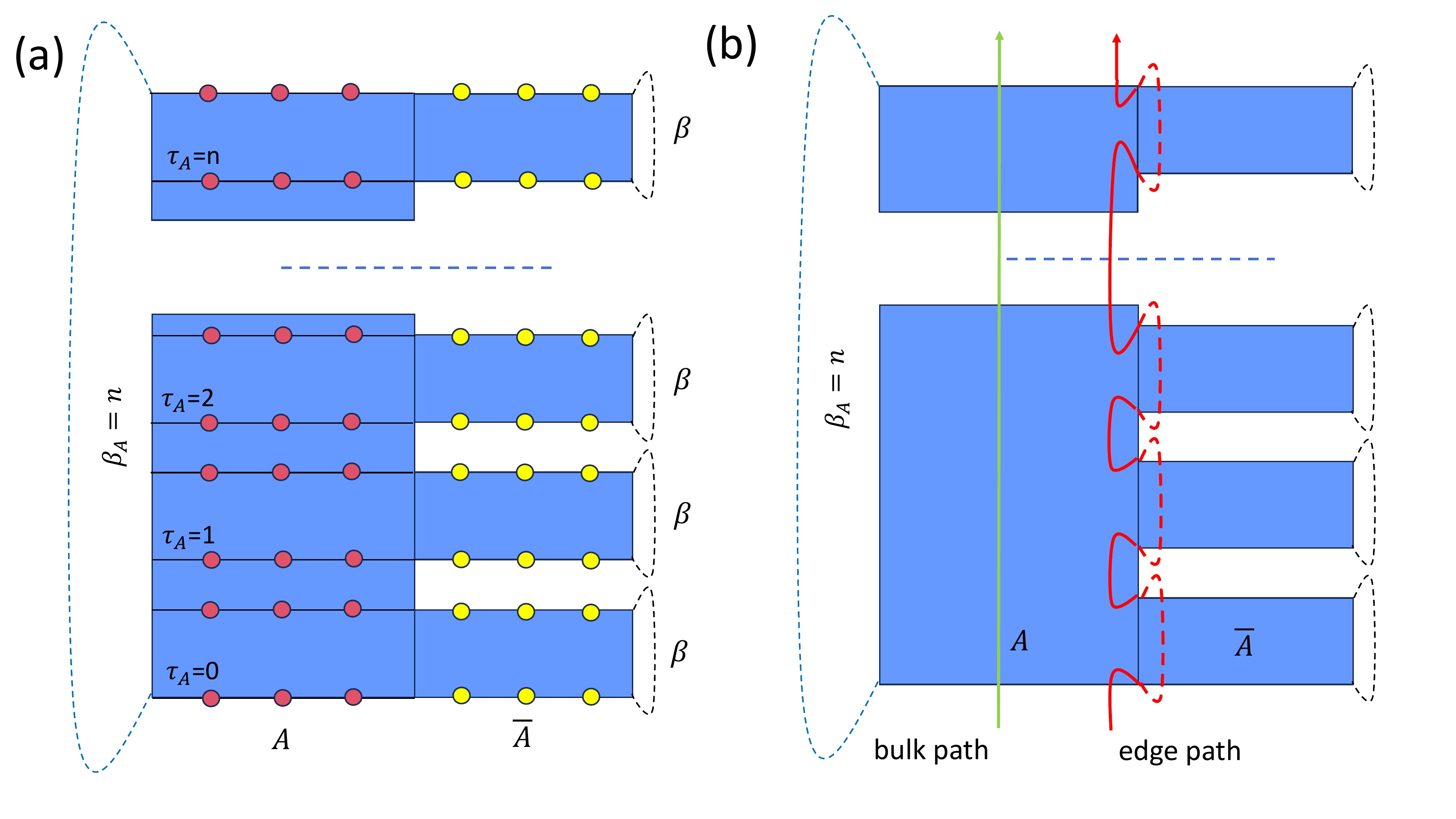}
 \caption{(a) A geometrical structure of the partition function $\mathcal{Z}^{(n)}_{A}$. The subsystem $A$ is entangled with the environment $\overline{A}$. Each replica connects with each other in $A$, while the environment $\overline{A}$ for each replica is independent. $\beta_A=n$ represents the imaginary-time length in $A$ and $\beta$ is $1/T$ for the total system. (b) A schematic picture of worldlines crossing a replica. The subsystem $A$ is coupled with the environment $\overline{A}$ via $J_{y}$. These worldlines which cross the bulk will decay to zero as $\beta \rightarrow \infty$ in principle. Meanwhile, the ones which cross the imaginary-time edge of $\overline{A}$ will arrive at the next replica without much cost. }
 \label{Fig.es-wormhole}
 \end{figure}

\begin{figure*}[htbp]
\centering
\includegraphics[width=0.85\textwidth]{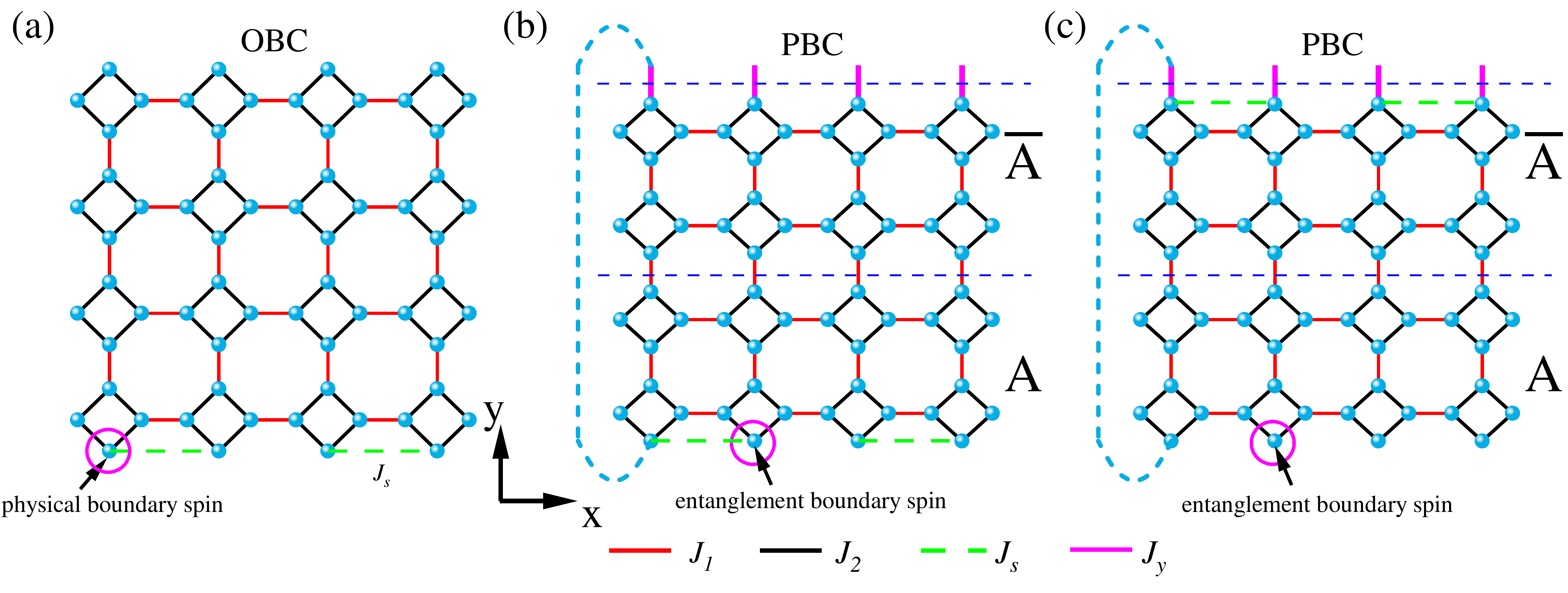}
\caption{Schematic figures of the model Eq.~\ref{eq2} on the square-octagon lattice under periodic boundary condition (PBC) along the $x$ direction and a tunable boundary in the $y$ direction. (a) Open boundary condition (OBC) in the $y$ direction and $J_s$ perturbations at the lower edge. (b-c) A tunable boundary coupling $J_y$ along the $y$ direction and $J_s$ perturbations at the entanglement boundary which is marked by blue dashed lines. $A$ and $\overline{A}$ denote the subsystem and environment respectively. The $J_y$ coupling can change boundary conditions along the $y$ direction. In particular, the system is under PBC or OBC when $J_y=1$ or $0$ respectively.}
\label{fig-lattice}
\end{figure*}

Numerical results have also shown that a relation between the low-lying ES and boundary energy spectrum exists generally in some magnetic systems beyond topological states~\cite{Poilblanc2010entanglement,zyan2021entanglement}. Then, it was theoretically shown to be a general relationship between the entanglement spectrum of $(2+1)$d chiral gapped topological states and the energy spectrum on their $(1+1)$d edges~\cite{XLQi2012}. Following studies~\cite{zhu2019reconstructing,Cirac2011,
Chiral,EntanglementAKLT} also confirm this conjecture. However, there is generally no exact correspondence between the low-energy spectrum of the edge Hamiltonian and entanglement Hamiltonian in the non-chiral topological phase without symmetry enrichment~\cite{Edge2015}, in which the boundary can be gapped.
A deep understanding of the correspondence between the entanglement and physical boundary spectrum is still lacking.
Even though the Li-Haldane conjecture is always satisfied in one-dimensional (1D) or quasi-1D systems which has been demonstrated widely in numerical results such as exact diagonalization (ED) and density matrix renormalization group (DMRG)\cite{Poilblanc2010entanglement,Pollmann2010entangle,Cirac2011}, whether it can hold in two-dimensional or higher dimensional systems is unclear so far.

Very recently, the extraordinary wormhole mechanism based on the replica manifold has been proposed to unlock the relation between the entanglement spectrum and energy spectrum \cite{zyan2021entanglement}. Via analysis of worldlines in the entangled edge and bulk (Fig.\ref{Fig.es-wormhole}), it successfully confirmed the edge mode held an important position in the ES, which explained the Li-Haldane conjecture well and provided a new window to understand the many-body entanglement spectrum. To probe the wormhole picture in a more general setting, we numerically study the entanglement properties of a two-dimensional Affleck-Kennedy-Lieb-Tasaki (AKLT) model with tunable boundaries. The typical positive and negative examples of the Li-Haldane conjecture can be realized to further explore the wormhole mechanism in this model.

In the wormhole picture, the coupling between the subsystem and environment makes the imaginary path length of imaginary time near the entangled boundary much shorter than that in the bulk, which makes the perturbation near the entanglement boundary in the environment important. As a result, the entanglement spectrum on the boundary may become different from the physical boundary energy spectrum, which is also consistent with the wormhole picture but beyond the Li-Haldane conjecture. The Li-Haldane conjecture can be regarded as a special case in the wormhole picture while all the examples we study can be self-consistent with the wormhole picture. Our paper suggests that the wormhole picture can be considered as a more universal physical picture to understand and predict the ES in general cases.

\section{Model and Method}
\subsection{A. Model}
In order to apply the wormhole effect picture to the perturbed boundary case and further demonstrate the general prediction beyond the Li-Haldane conjecture, we investigate the $S=1/2$ Heisenberg model on the square-octagon lattice with $C_{4v}$ lattice symmetry, whose Hamiltonian in the bulk can be written as
\begin{equation}
\begin{split}
H_b = J_{1}\sum_{\langle ij\rangle}\mathbf{S}_{i}\cdot \mathbf{S}_{j}+J_{2}\sum_{\langle ij\rangle'}\mathbf{S}_{i}\cdot \mathbf{S}_{j}
\end{split}
\end{equation}
where the inter-unit-cell coupling $J_{1}=1$ as the unit energy scale and intra-unit-cell coupling $J_{2}>0$. The model hosts a rich phase diagram for the ground states under the competition between $J_1$ and $J_2$ terms.
There are four phases including the $S=2$ N\'eel phase, AKLT phase, $S=1/2$ N\'eel phase and plaquette valence bond crystal, which are separated by three $O(3)$ quantum critical points~\cite{PhysRevLett.118.087201}. The AKLT state is a symmetry protected topological (SPT) state~\cite{AKLT,spt2009} and has gapless boundaries protected by the translation invariance and spin rotation symmetry. The gapless boundary can be modeled by an effective $S=1/2$ Heisenberg chain~\cite{PhysRevB.105.014418}. This model is the simplest two-dimensional SPT model and free of sign problems, which is easy to implement in quantum Monte Carlo (QMC) \cite{Sandvik1999,sandvik2010computational}. What is more, the AKLT phase is a typical SPT phase with bulk-boundary correspondence and can be well described by the Li-Haldane conjecture. Surprisingly, the influence of the gapless boundary even extends the physics of the $O(3)$ critical point and induces unconventional surface critical behaviors~\cite{PhysRevLett.118.087201}.  To investigate the wormhole effect picture deeply, we add other terms to the original Hamiltonian to make the system far away from the standard Li-Haldane conjecture case:
\begin{equation}
\begin{split}
\label{eq2}
H = H_b + J_{s}\sum_{\langle ij\rangle_{s}}\mathbf{S}_{i}\cdot \mathbf{S}_{j} +J_{y}\sum_{\langle ij\rangle_{y}}\mathbf{S}_{i}\cdot \mathbf{S}_{j}
\end{split}
\end{equation}
where $J_s$ is a small translation-invariance-breaking term on the boundary and $J_y$ is the coupling between the subsystem $A$ and environment $\overline{A}$ which can change the boundary condition along the $y$ direction.  Fig.~\ref{fig-lattice} depicts the model Eq.~\ref{eq2} on the square-octagon lattice. By tuning the perturbation and coupling on the edge, positive and negative examples of the Li-Haldane conjecture can be all explored well.

\begin{figure*}[htb]
\centering
\includegraphics[width=0.90\textwidth]{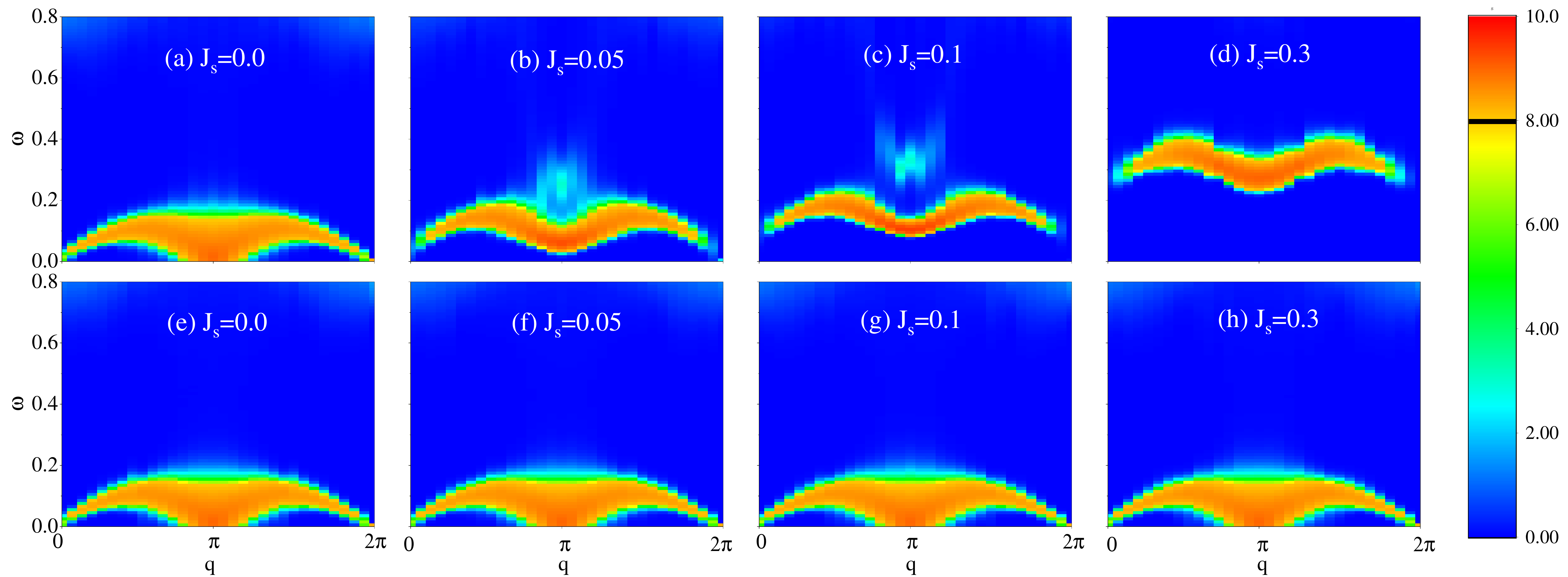}
\caption{Dynamical spectra for the $J_1-J_2$ Heisenberg model on the square-octagon lattice under a modified boundary as in Fig.~\ref{fig-lattice}(a) with $L=32$ and $\beta=64$. Upper row: Dynamical spectra on the bottom boundary. Lower row: Dynamical spectra on the top boundary. For better presentation, we set the color bars to be logarithmic scale above $8.0$.}
\label{Fig.es3}
\end{figure*}

\begin{figure*}[htb]
\centering
\includegraphics[width=0.90\textwidth]{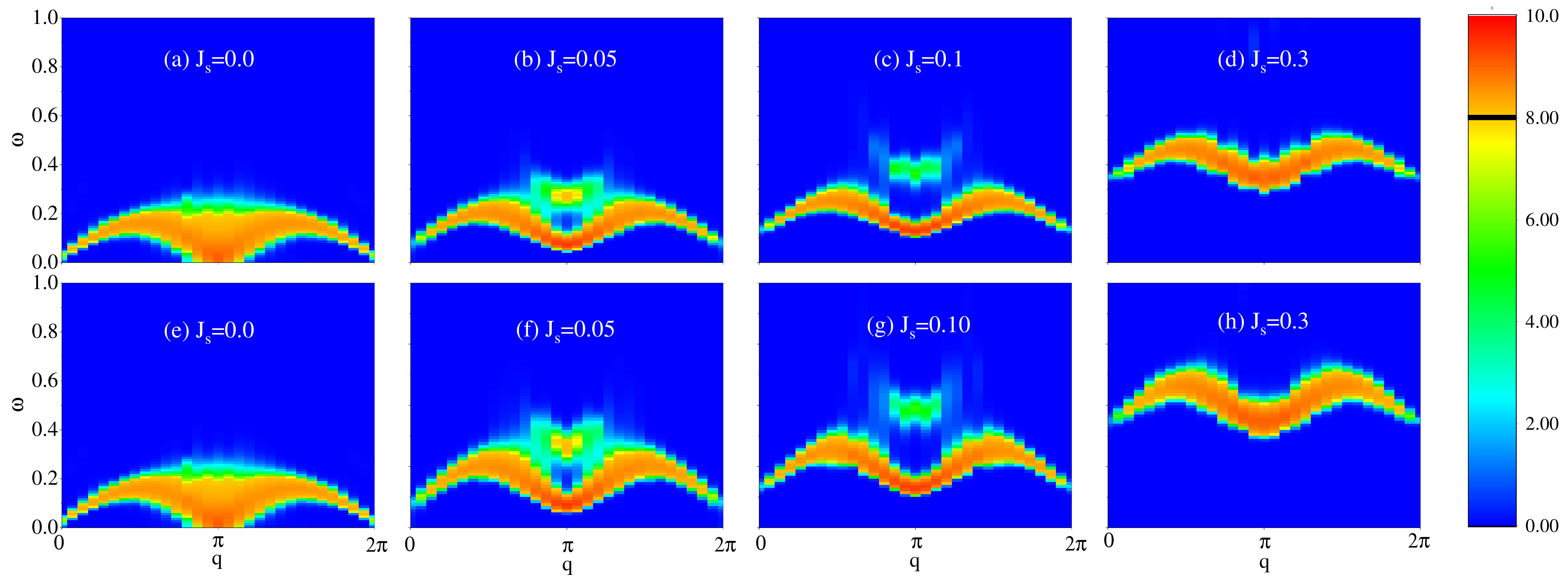}
\caption{Entanglement spectra for the $J_1-J_2$ Heisenberg model on the square-octagon lattice under a modified entanglement boundary as in Fig.\ref{fig-lattice} (b) and (c) with $L=32$, $\beta_A=64$ and $\beta=50$. Upper row: Entanglement spectra of the bottom edge as shown in Fig.\ref{fig-lattice} (b). Lower row: Entanglement spectra of the bottom edge as shown in Fig.\ref{fig-lattice} (c). For better presentation, we set the color bars to be logarithmic scale above $8.0$.}
\label{Fig.es4}
\end{figure*}

\subsection{B. Method}
We use the recently proposed QMC-based numerical method to study the low-lying ES~\cite{zyan2021entanglement,wu2023classical}. In the framework of QMC, it is difficult to construct the $\rho_{A}$=$Tr_{\overline{A}}(|\psi \rangle \langle\psi|)=e^{-\mathcal{H}_{A}}$ directly as ED~\cite{Poilblanc2010entanglement} and DMRG do~\cite{Cirac2011,Chiral}. The ES can be obtained through stochastic analytic continuation (SAC) from the imaginary-time evolution of $\rho_{A}$ in the QMC simulations~\cite{Constrained2016,Nearly2017,SHAO2023}. Specifically, by introducing an effective imaginary time $n$ for the entanglement Hamiltonian, the effective partition function can be written as
\begin{equation}
\begin{split}
\label{eq3}
\mathcal{Z}^{(n)}_{A}=Tr[\rho^{n}_{A}]=Tr[e^{-n \mathcal{H}_{A}}]
\end{split}
\end{equation}
where the effective imaginary time $\beta_{A}=n$. By constructing the modified manifold in the QMC simulations (Fig.~\ref{Fig.es-wormhole}), we can obtain the imaginary-time spin correlation $G(\tau_{A},\mathbf{q})=\langle S^{z}_{-\mathbf{q}}(\tau_{A})S^{z}_{\mathbf{q}}(0)\rangle$ for the entanglement Hamiltonian $\mathcal{H}_{A}$ and then find the corresponding ES in the spectral function $G(\omega,\mathbf{q})$ via SAC. 
We mainly concentrate on the AKLT state of the model (Eq.~\ref{eq2}) in the bulk with the system size $32\times 32$ and $J_2=0.3$. The bulk has a large energy gap, where the low-energy physics is governed by its boundary. In addition, due to the spin rotation symmetry of the model, we can only consider the imaginary-time boundary correlation for the $s^z$ operator, $G(\tau_{A},\mathbf{q})=\frac{1}{L}\sum_{i,j}e^{-i\mathbf{q}_{x}\cdot(\mathbf{x}_{i}-\mathbf{x}_{j})}\langle s^{z}_{i}(\tau)s^{z}_{j}(0)\rangle$, where $s^z_{i}$ denotes spins on the physical or entanglement boundary.

\subsection{C. The wormhole picture}
The entanglement spectrum algorithm method inspired an interesting wormhole effect inducing the low-lying ES to uncover the mysteries of entanglement spectrum~\cite{zyan2021entanglement}. In the replica manifold as Fig.~\ref{Fig.es-wormhole} (b) shows,
the trace of environment connects the upper and lower imaginary-time boundaries of one replica, so it provides a convenient way for worldlines to get to the upper edge from the lower edge instantly, dubbed the wormhole picture. It makes the worldline avoid going through the whole bulk replica with a huge cost. As Fig.\ref{Fig.es-wormhole} shows, the cost of the path in the path integral is proportional to $\overline{\mathcal{L}} \times \Delta(\overline{\mathcal{L}})$, where $\overline{\mathcal{L}}$ is the average path length of imaginary time and $\Delta(\overline{\mathcal{L}})$ is the average gap for this path. The smaller cost $\overline{\mathcal{L}} \times \Delta(\overline{\mathcal{L}})$ will lead to the larger path weight which holds a more important role in the low-lying spectrum. Via analysis of the paths in the bulk and near the entangled edge at Fig.\ref{Fig.es-wormhole}, the bulk path length should be proportional to $\beta \times n$ and edge path length should be proportional to $1 \times n$. Then, the ratio of path length can be regarded as $\beta:1$.\\

According to comprehensive  consideration of path length and gap, the ratio of cost between bulk path and edge path can be estimated roughly as $\beta \Delta_b : \Delta_e$, where $\Delta_b$ is the bulk gap and $\Delta_e$ is the edge gap. The inverse temperature $\beta \rightarrow \infty$ at the ground state which renders the $\beta \Delta_b \gg \Delta_e$. Therefore, the cost of the edge path near the entangled boundary  (red line of Fig.~\ref{Fig.es-wormhole} (b)) is much lower than in the bulk (green line of Fig.~\ref{Fig.es-wormhole} (b)). Because the lower cost of the path provides larger weight in the path integral, the edge mode plays an important role in the low-lying ES. Thus the Li-Haldane conjecture has been explained well within the wormhole mechanism.\\ 

However, if there is no bulk-boundary correspondence, the general relation between entanglement spectrum and energy spectrum is not clear so far.  Generally, when the boundary is influenced by perturbation, the change of ES is an open question which is not guaranteed to be like the energy spectrum under the Li-Haldane conjecture. According to the wormhole effect mechanism, how the worldline crosses the edge is a key point to unlock the relation between the entanglement spectrum and energy spectrum. Therefore, it is interesting to apply the wormhole effect picture to the perturbed boundary and predict the entanglement spectrum.

\section{Results}
In this section, we first study the influence of the perturbation term $J_s$ on the entangled edge to the ES with a full coupling strength $J_y=1$, i.e., under the periodic boundary condition (PBC) in the $y$ direction. Then we tune the coupling $J_y$ to systematically explore how the correspondence between the boundary ES and energy spectrum breaks down, including the case of $J_y=0$ i.e. the case of open boundary condition (OBC).

\subsection{A. Probing the perturbed boundaries}
As Fig.~\ref{fig-lattice} illustrates, when the boundary is gapped by the perturbation, we make a comparison study between low-lying spectra on the physical and entanglement boundaries, i.e., the energy spectra on the physical boundary for a system under OBC and entanglement spectra on the entanglement boundary for a system under PBC ($J_y = 1$).

The energy spectra are presented in Fig.~\ref{Fig.es3} through doing numerical analytic continuation for the boundary spin-spin dynamical spectral function. Without the translation-invariance-breaking term, $J_s=0$, two physical boundaries are both gapless and can be modeled by a one-dimensional spin-$1/2$ Heisenberg model. The spin-spin dynamical spectral function corresponds to the two-spinon excitation, which is consistent with the feature of the gapless boundary of SPT phases. Without symmetry protection, the boundary is not guaranteed to be gapless. As shown in Fig.~\ref{Fig.es3} (a)-(d), when $J_s$ increases, the perturbed boundary becomes gapped in which the excitation becomes a magnon and the energy gap is larger for stronger $J_s$. Meanwhile, the other untouched boundary remains gapless (see Fig.~\ref{Fig.es3}(e)-(h)). This case is beyond Li-Haldane's condition, and we will recount below that the ES can still be explained by the wormhole mechanism while the Li-Haldane conjecture fails.

Interestingly, for the entanglement case (not really an open edge but replaced by an entangled edge), we find the entanglement spectrum in Fig.~\ref{Fig.es4} does not always show similar features as the energy spectrum of subsystem $A$ as an independent physical system under OBC. Without the perturbation term, $J_s=0$, the entanglement boundary is gapless as demonstrated in Fig.~\ref{Fig.es4}(a) and (e), which is consistent with the Li-Haldane conjecture.

 With a small $J_s$ term on the entanglement boundary, the entanglement spectrum on the lower boundary of $A$ becomes gapped, whether we tune the boundary coupling $J_s$ in $A$ or $\overline{A}$ as shown in Fig.~\ref{Fig.es4} (b)-(d) and Fig.~\ref{Fig.es4} (f)-(h) respectively.
Even if the case of tuning the $J_s$ in $A$ barely satisfies the Li-Haldane conjecture that the ES opens a gap similar as the edge energy spectrum does, it can be treated as an extended conjecture. However, the case of tuning the $J_s$ in $\overline{A}$ totally violates the prediction of similar behaviors between the ES and edge energy spectrum, because in this case the edge energy spectrum of $A$ is still gapless due to the fact that $J_s$ is tuned in environment $\overline{A}$. But we observe that the ES opens a gap.
\begin{figure*}[htbp]
\centering
\includegraphics[width=0.90\textwidth]{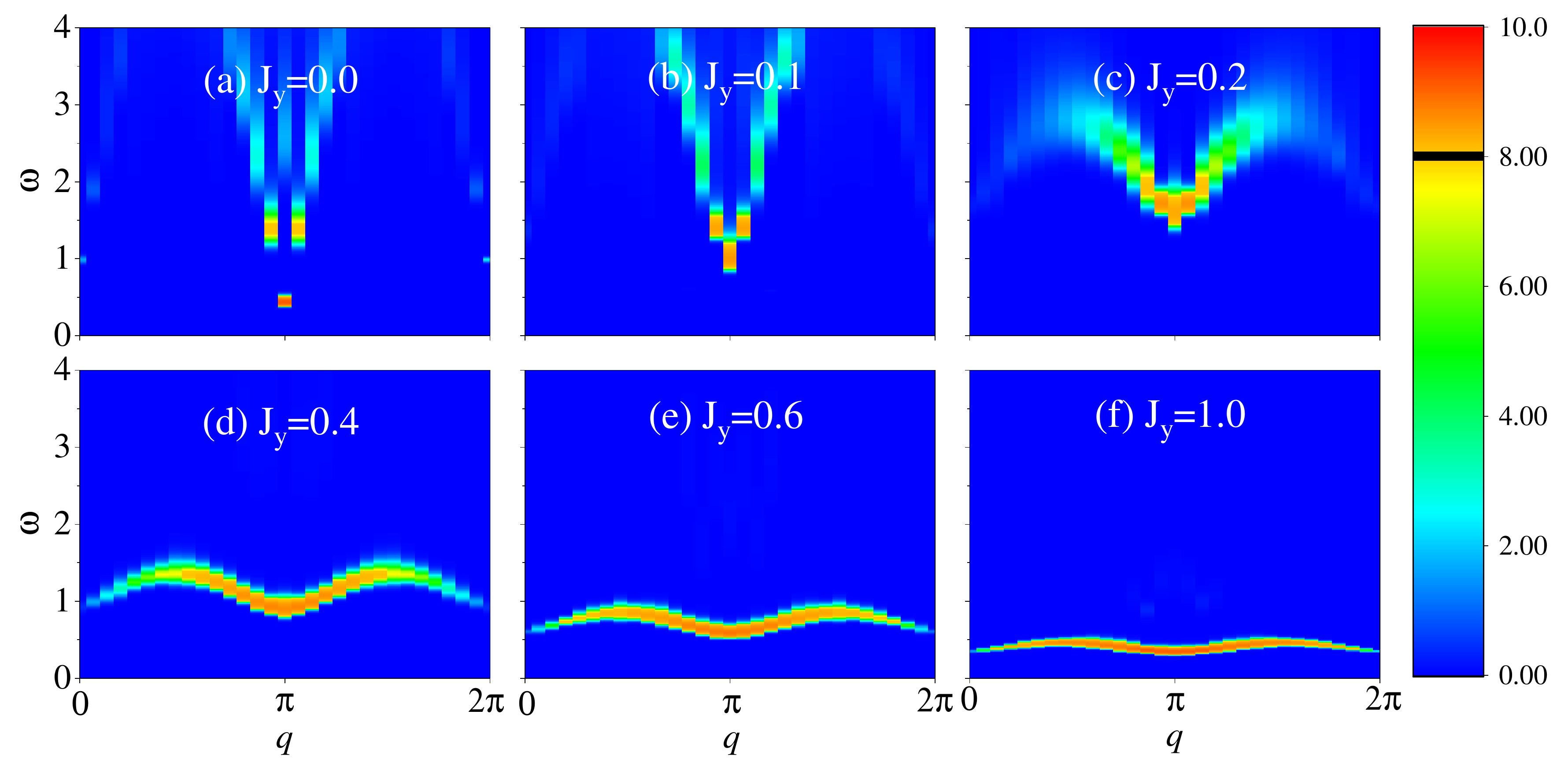}
\caption{The entanglement spectra of Fig.\ref{fig-lattice} (c) with $J_{s}=0.3$, $L=32$, $\beta=50$ and  $\beta_A=64$, where $J_{y}$ is (a)-(f) 0.0, 0.1, 0.2, 0.4, 0.6, and 1.0. For better presentation, we set the color bars to be logarithmic scale above 8.0.}
\label{Fig.es4-2}
\end{figure*}

Actually, according to the path-integral wormhole picture for the ES, this change can be understood naturally~\cite{zyan2021entanglement}. As shown in Fig.\ref{Fig.es-wormhole} (b), the red line path near the entangled edge is the shortest path to contribute to the low-lying ES. Under the PBC, or with a full strength coupling $J_y=1$, the path goes through the edges of both $A$ and $\overline{A}$ alternately, thus the perturbation on the lower boundary of $A$ or upper boundary of $\overline{A}$ plays a similar role in the path integral and both of them results in the gapped boundary ES. In the proposed wormhole mechanism picture (Fig.~\ref{Fig.es-wormhole} (b))~\cite{zyan2021entanglement}, the ratio of the cost between the bulk path (green line) and edge path (red line) can be approximately regarded as $\beta \Delta_{b}:\Delta_{e}$, where $\Delta_{b}$ and $\Delta_{e}$ refer to the energy gap of the bulk and edge.  For our model, the inverse temperature $\beta$ should go to $\infty$ in the bulk so the low-energy entanglement spectrum should look similar to the generalized entanglement edge of the energy spectrum whether or not the perturbation gaps the generalized entanglement edge.   It clearly shows that the limitation of the Li-Haldane conjecture can be detected via the tunable perturbation, while the wormhole picture can give the self-consistent explanation for ES in the gapless or gapped cases. Besides, if we gap the lower edge of $A$ and the upper edge of $\overline{A}$ via adding the perturbation $J_s$ on both entanglement boundaries of $A$ and $\overline{A}$, the ES gap becomes twice as large as the case that only one boundary is gapped (see Appendix C).

\subsection{B. The role of the coupling between system and environment} 

To further explore the influence of wormhole effect on the correspondence between energy and entanglement spectrum, we systematically study the ES with different coupling strength $J_y$ between the subsystem and environment.

The inverse temperature $\beta$ is $50$ and $\beta_A$ is $64$ with the environment perturbation $J_s=0.3$ (as Fig.\ref{fig-lattice}(c) shows). When $J_y$ is from $0$ to $0.2$, the ES gap gradually becomes large. And then the gap gets smaller and smaller with increasing the coupling strength as Fig.\ref{Fig.es4-2} shows. When $0<J_y<1$, the ES becomes gapped which is neither like the gapless boundary of system $A$ nor like the full gapped boundary of the environment of $\overline{A}$ (the ES gap is different). The gap change looks non-monotonous which is different from the $J_s=0$ case with only wormhole effect (see Appendix A). 
According to the wormhole picture, the cost of the edge path should be proportional to $\Delta_e/f(J_y)$, where $f(J_y)$ is supposed to be a monotonically increasing polynomial function of the coupling $J_y$ (When $J_y$ is small, $f(J_y)$ is very close to zero). Then the ratio of cost between bulk path and edge path can be roughly corrected as $\beta \Delta_b:\Delta_e/f(J_y)$. When $J_y$ is weak near zero, the edge cost  will become very large beyond $\beta \Delta_b$ so that more worldlines will choose the bulk path. Increasing $J_y$ will transport more environment perturbation to the subsystem, which makes the bulk gap $\Delta_b$ larger. Therefore, the ES gap becomes large as we increase $J_y$ as shown in Fig.\ref{Fig.es4-2} (a)-(c).


For strong $J_y$, the cost of the edge path $\Delta_e/f(J_y)$ becomes less than $\beta \Delta_b$ due to the large $f(J_y)$, which leads more worldlines to go through the edge path.  More importantly, the wormhole effect can hold the dominant position when $J_y$ is large enough. It makes the ES gap becomes smaller as we continue increasing the coupling to the full strength $J_y=1$, which is finally like the energy spectrum of the environment boundary (Fig.\ref{Fig.es4-2}(d)-(f)).   

When tuning the coupling, the correspondence between entanglement spectrum and energy spectrum becomes invalid which predicts the gapless ES inconsistent with numerical results. The ES on the boundary of gapped phases can be changed via the modification of the environment and the coupling between the subsystem and environment. The wormhole picture explains why the ES can be hugely influenced by the coupling. This analysis reveals that the interplay between the coupling and perturbation is also well captured by the wormhole picture with replica amplification of the bulk gap, which is a powerful tool to understand the entanglement spectrum of systems with more complex interaction and perturbation.

\section{DISCUSSION}
In the original Li-Haldane conjecture, the coupling strength between subsystem and environment was not explored. It is usually limited to special systems that have SPT order or chiral topological order. It has been shown that such a correspondence may not hold in the non-chiral topological ordered states. There is no general physical picture to explain it. Our paper reveals that an important missing point is the role of coupling between the subsystem and its environment. From the path-integral (wormhole) picture for the partition function in the replica manifold (Fig.~\ref{Fig.es-wormhole}), it is clear that the coupling makes the boundary of $A$ and $\overline{A}$ contribute much more importantly than the bulk. The worldlines near the entanglement boundary cross the edge of $A$ and $\overline{A}$ alternately so that both boundary perturbations of $A$ and $\overline{A}$ contribute to the low-lying ES. This wormhole effect provides a simple picture and a fundamental mechanism for the inconsistency of correspondence between the energy and entanglement spectrum on the boundary.

In this paper, we limited our study near the boundaries for a gapped SPT phase and symmetric couplings $J_y$ between the subsystem and its environment. It is not clear how the entanglement spectrum will be changed by perturbations and couplings in general cases. It would be still natural that the coupling between the system and environment also serves as a key role. Without the coupling between the subsystem and its environment, local perturbations in the environment have large imaginary time in the replica manifold and can not contribute a lot to the ES. Although we only concentrated on one boundary, these analyses also apply to the other unperturbed boundary, i.e, the upper or lower boundary of $A$ or $\overline{A}$ respectively. The importance of the coupling between the subsystem and its environment in the wormhole picture also reflects the quantum nature of entanglement. Without any coupling, the whole system becomes a tensor product of $A$ and $\overline{A}$ and there would be no entanglement at all. And for no coupling, the entanglement boundary of subsystem $A$ can be naturally described by an effective $S=1/2$ Heisenberg chain (see Appendix D). When the bulk is a critical point or gapless phase, it is still not clear whether the wormhole picture applies. In the wormhole picture, the ratio of cost between the bulk path and edge path is approximately $\beta \Delta_b:\Delta_e$. As we know, $\Delta_b$ and $\Delta_e$ are zero at the quantum critical point or gapless phase. Both bulk mode and edge mode may cross the replica with similar cost, which are mixed together to contribute to the low-lying ES.  Thus, it is more complex for analyzing the contributions from bulk and edge for the low-lying ES at the  quantum critical point or gapless phase.

\section{CONCLUSION}
In summary, we have systematically studied the relation between the energy and entanglement spectrum on the boundary in the AKLT phase of the $S=1/2$ Heisenberg model on the square-octagon lattice with a symmetry-breaking term on the boundary and a tunable coupling between the subsystem and its environment. The correspondence does not always apply if the perturbation is added to the environment. The entanglement spectrum can show different behaviors from the energy spectrum of the subsystem as an independent system under open boundary condition. Tuning the coupling between the subsystem and environment can change the ES gap, which indicates that the coupling accounts for the inconsistency of the Li-Haldane conjecture. This inconsistency can be naturally explained in the wormhole picture for the partition function in the replica manifold. In the path-integral picture, the wormhole effect induced by the coupling plays an important role in transporting the interaction of the environment and provides a pathway to control the ES. The wormhole picture can explain the correspondence between the entanglement spectrum and energy spectrum for the gapless and gapped cases which is indeed the fundamental mechanism of the ES.
The rapidly developed quantum circuits experiments have successfully extracted the ES through measuring different orders of R\'enyi entropies~\cite{Johri2017Entanglement,Chang2019Evolution,Zhang2020Nonuniversal,lavasani2021measurement}. The basic idea is to fit the low-lying ES according to the $1\sim n$ order R\'enyi entropies~\cite{JRZhao2021}, where the $n$ is a cut-off order. Besides, the recent cold atom experiment has successfully probed the Li-Haldane conjecture in the topological states by using entanglement Hamiltonian tomography and quantum variational learning ~\cite{Zache2022entanglement}. It makes the possibility that
our results can be realized in quantum circuits and cold atom systems via constructing and controlling the AKLT or other SPT states.

\section{Acknowledgements}
Z.Y. would like to thank Bin-Bin Chen, Shangqiang Ning and Zi Yang Meng for fruitful discussions. D.X.Y. and Z.L. are supported by NKRDPC-2022YFA1402802, NKRDPC-2018YFA0306001, NSFC-11974432, NSFC-92165204, GBABRF-2019A1515011048, Leading Talent Program of Guangdong Special Projects (201626003), and Shenzhen International Quantum Academy (Grant No. SIQA202102). The authors acknowledge Beijng PARATERA Tech Co.,Ltd. for providing HPC resources that have contributed to the research results reported within this paper. R.Z.H. is supported by a postdoctoral fellowship from Ghent university - Special Research Fund (BOF). Z.Y. thanks the support from the start-up fund of Westlake University and the open fund of Lanzhou Center for Theoretical Physics (12247101). Z. L. acknowledges a one-month hospitality of Westlake University invited by Z.Y..
%

\newpage
\appendix

\section{Appendix A: Different coupling strength with $J_s=0.0$}

\begin{figure}[htp]
\centering
\includegraphics[width=0.50\textwidth]{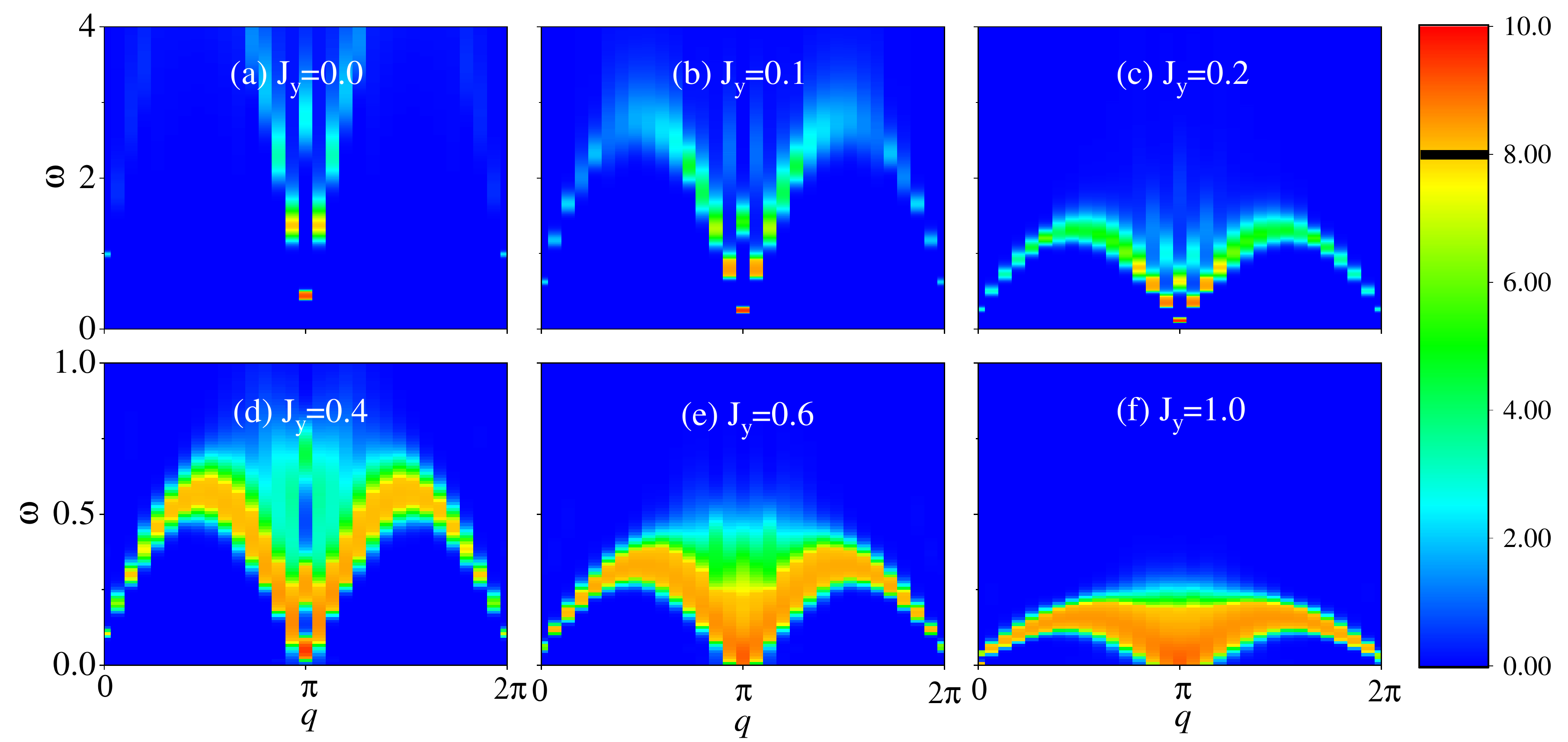}
\caption{The original entanglement spectra of Fig.\ref{fig-lattice} (c) with $J_{s}=0.0$, $L=32$, $\beta=50$ and $\beta_A=64$, where $J_{y}$ is (a)-(f) 0.0, 0.1, 0.2, 0.4, 0.6, and 1.0. For better presentation, we set the color bars to be logarithmic scale above $8.0$. }
\label{Fig.es-js0-1}
\end{figure}

In this section, we tune $J_{y}$ from $0$ to $1.0$ with $\beta=50$ and $\beta_{A}=64$ to explore the wormhole effect on ES without normalization. The wormhole effect can lower the edge modes which is clearly shown in the original ES with $J_s=0.0$ in Fig.\ref{Fig.es-js0-1}. Because $\Delta \tau=1$ for replica structure is too large to obtain the high-energy part on ES by SAC, we only observe the low-energy excitation in Fig.\ref{Fig.es-js0-1} (a) and (b), which is enough to analyze the important entanglement message. At the $J_y=0$ limit, the edge becomes a real physical edge and the ES has a small gap. Although the ES should become a gapless two spinon continuum in the AKLT phase, the finite-size gap becomes larger in ES because the wormhole effect disappears at $J_y=0$. The gap is amplified $\beta$ times due to the replica system. This amplification effect leads to the ES gap.

Then, the gap of the ES becomes smaller and smaller when $J_y$ increases, which demonstrates that the wormhole effect is magnified as $J_y$ increases. Thus, the wormhole effect makes the ES gap become smaller as $J_y$ increases, which lowers the bulk amplification effect and makes the bandwidth become smaller at the same time. Actually, the gap change can not be derived directly from the original Li-Haldane conjecture. The coupling strength plays an important role in the ES mechanism which has been confirmed in our numerical results.

\section{Appendix B: More boundary interactions}

\begin{figure*}[htbp!]
 \centering
 \includegraphics[width=0.93\textwidth]{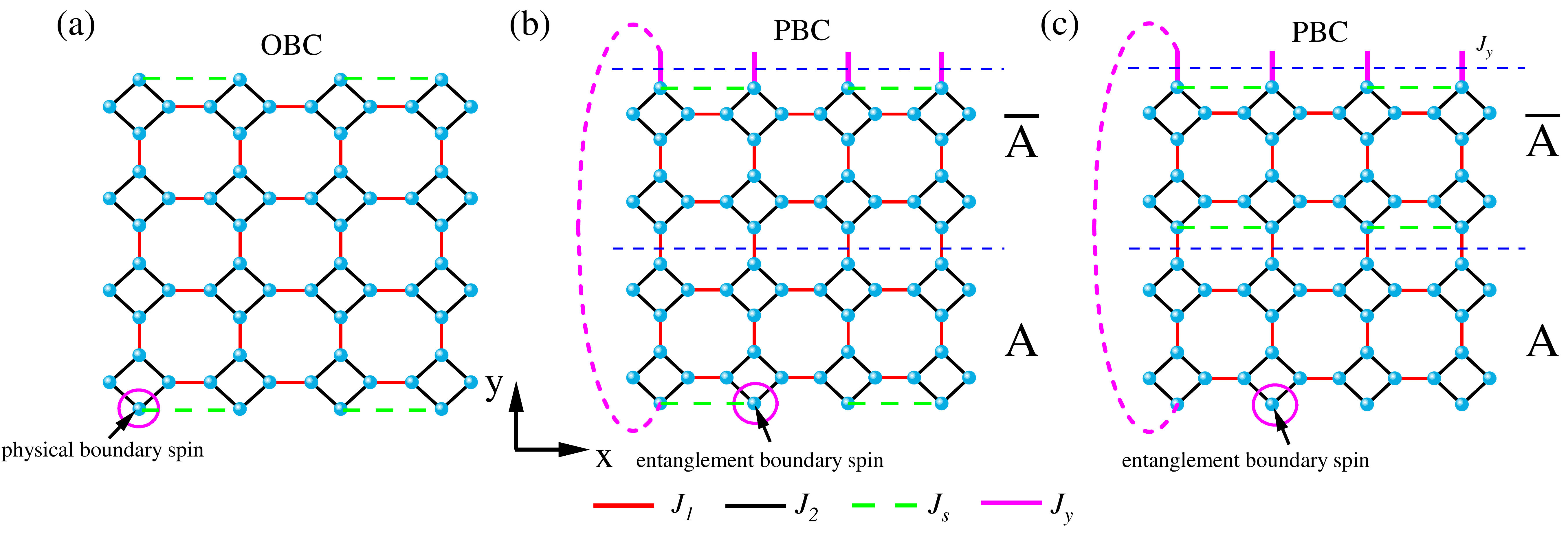}
 \caption{The square-octagon lattice with the PBC in $x$ direction. (a) The lattice with OBC in $y$ direction with $J_{y}=1$. (b) The lattice with PBC in $y$ direction. The blue dashed line cuts it into two entangled parts, $A$ and $\overline{A}$. And $J_{s}$ is added to the lower boundary of subsystem $A$ and upper boundary of $\overline{A}$ as green dashed lines show . (c) The lattice with PBC in $y$ direction, where $J_{s}$ is added to the upper boundary and lower boundary of environment $\overline{A}$ as green dashed lines show. }
 \label{fig-lattice-1}
 \end{figure*}

\begin{figure}[htbp!]
 \centering
 \includegraphics[width=0.48\textwidth]{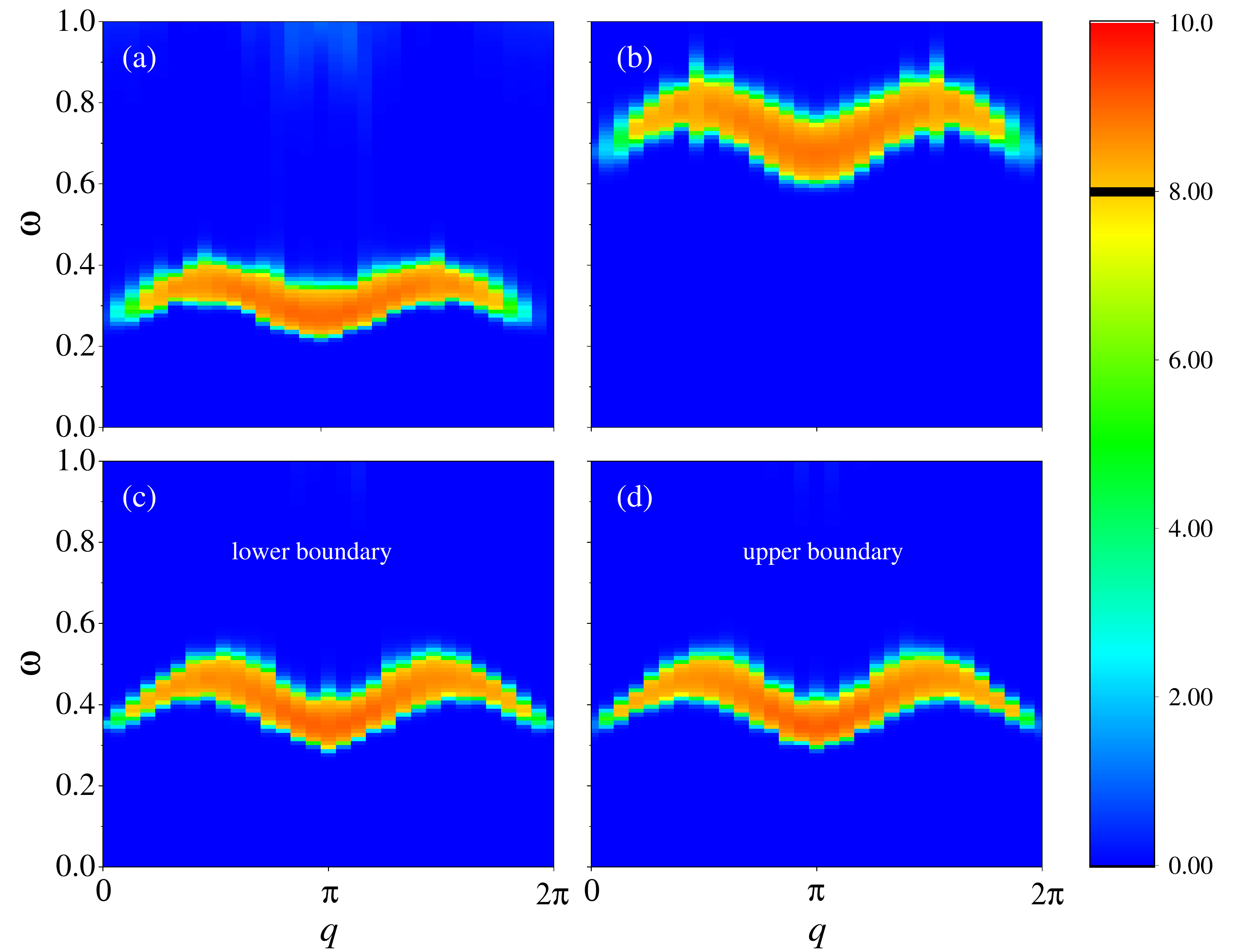}
 \caption{The original dynamical spectra ($\beta=64$) and entanglement spectra ($\beta=50$ and $\beta_A=64$) of Fig.\ref{fig-lattice-1} with $J_{y}=1$ and $L=32$. (a)Dynamical spectrum with $J_{s}=0.3$. (b) Entanglement spectrum of the lower boundary of subsystem $A$ at Fig.   \ref{fig-lattice-1} (b) with $J_{s}=0.3$. (c) Entanglement spectrum of the lower boundary of subsystem $A$ at Fig.\ref{fig-lattice-1} (c) with $J_{s}=0.3$. (d) Entanglement spectrum of the upper boundary of subsystem $A$ at Fig.\ref{fig-lattice-1} (c) with $J_{s}=0.3$. For better presentation, we set the color bars to be logarithmic scale above $8.0$.}
 \label{Fig.es5}
 \end{figure}

If more boundary interactions are tuned near the entanglement edge, it is unclear whether the ES will be like the edge spectrum. We focus on the energy spectrum and entanglement spectrum of Fig.\ref{fig-lattice-1}, where the perturbation $J_{s}$ is added to two boundaries with $J_y=1$. When $J_{s}=0.3$, the edge spectrum of the model becomes gapped excitation (Fig.\ref{Fig.es5} (a)) which is similar to Fig.\ref{Fig.es3} (d). If the perturbation $J_s$ is added to the bottom edge of the subsystem and top edge of the environment to gap them at the same time (Fig.\ref{fig-lattice-1} (b)), the ES becomes also gapped (Fig.\ref{Fig.es5} (b)). And its gap is nearly twice as large as the ES in Fig.\ref{Fig.es5} (c-d), which demonstrates that both subsystem and environment perturbation contribute to the ES equally. It further proves that the coupling $J_y$ can transport the environment perturbation to the subsystem near the entanglement boundary. Furthermore, we simulate the ES with $J_{s}=0.3$ on the lower and upper boundaries of environment $\overline{A}$. In our expectation, the ES is also gapped on  both boundaries of subsystem $A$ as Fig.\ref{Fig.es5}(c) and (d) shows, which are similar to the energy spectrum.

\section{Appendix C: boundary correlation function}

 \begin{figure}[htbp]
 \centering
 \includegraphics[width=0.46\textwidth]{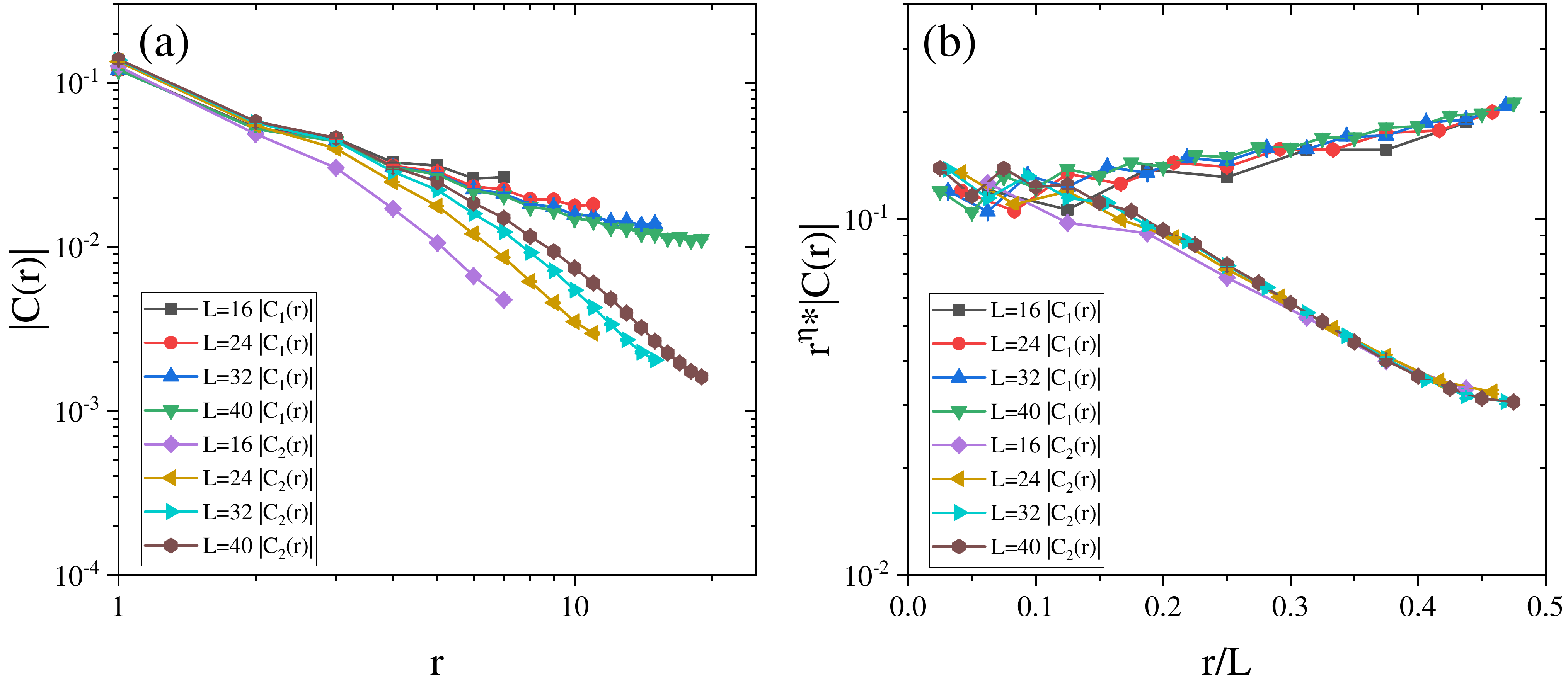}
 \vspace{5pt}
 \includegraphics[width=0.30\textwidth]{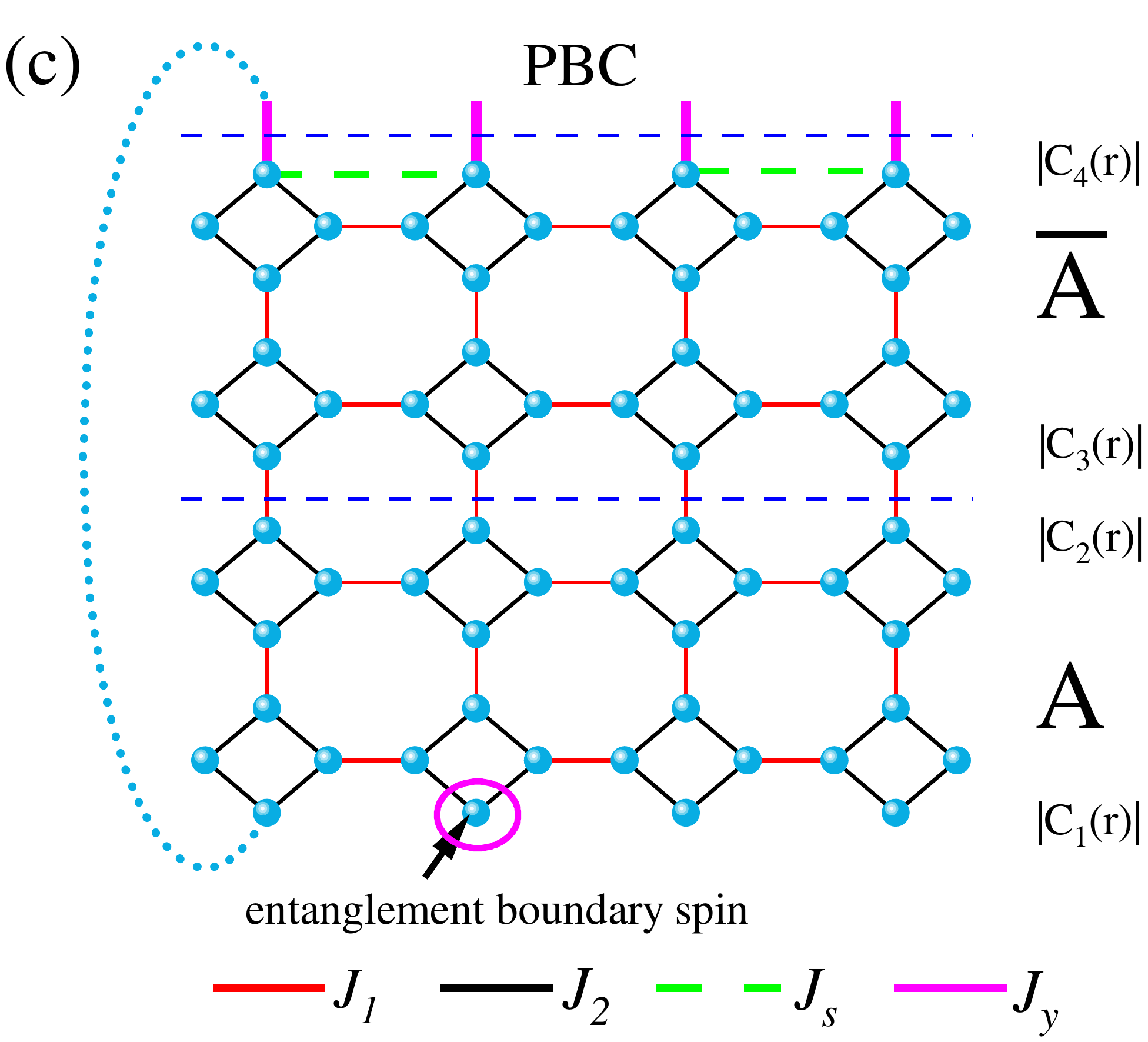}
 \caption{Different lattice sizes $L$ of boundary equal-time correlation functions $C_{k}(r)$ at $J_{s}=0.3$, $J_{y}=0.0$,  $\beta_A=2L$ and $\beta=50$. (a) The original boundary equal-time correlation $\lvert C_{1}(r) \rvert $ and $\lvert C_{2}(r) \rvert $ with $L$=16, 24, 32 and 40. (b) Rescaling of $\lvert C_{1}(r) \rvert $ and $\lvert C_{2}(r) \rvert $ for different sizes $L$. The exponent $\eta=1$ is the anomalous dimension. (c) $C_{k}(r)$ is marked at different entangled edges on the square-octagon lattice ($k$=1,2,3 and 4).}
 \label{Fig.es-js03-cr3}
 \end{figure}


In order to detect the feature of the entanglement Hamiltonian, we obtain different sizes of the boundary equal-time correlation functions $\lvert C_{1}(r) \rvert$ and $\lvert C_{2}(r) \rvert$ in Fig.\ref{Fig.es-js03-cr3}, where $C_{k}(r)=\langle S^{z}_{i} S^{z}_{j}\rangle$ marked in Fig.\ref{Fig.es-js03-cr3} (c) . The system sizes in the simulation are from $16$ to $40$ with $\beta=50$ and $\beta_A=2L$. $\lvert C_{1}(r) \rvert$ represents the behaviors for the entanglement boundary of $A$ without coupling, while $\lvert C_{2}(r) \rvert$ shows the behaviors for the boundary of $A$ with the full strength coupling. As shown in Fig.\ref{Fig.es-js03-cr3} (3), it is clear that $\lvert C_{1}(r) \rvert$ and $\lvert C_{2}(r) \rvert$ decay in a power law which is consistent with the gapless boundary behaviors.  This means that the bound of $A$ without coupling is similar to the bound of $A$ with full-strength coupling. After rescaling the correlation functions $C(r)=r^{-\eta}f(r/L)$ with $\eta=1$, we show that $\lvert C_{1}(r) \rvert$ and $\lvert C_{2}(r) \rvert$ can almost collapse into a single curve. The exponent $\eta=1$ is the anomalous dimension which is suitable for the (1+1)D gapless boundary of the AKLT state, though there is a logarithm correction to the correlation function $C(r) \propto 1/r$ in a $S=1/2$ Heisenberg chain. And the collapse picture clearly shows that $\lvert C_{1}(r) \rvert$ and $\lvert C_{2}(r) \rvert$ decay with the same power-law exponents, which indicates the entanglement Hamiltonian can be modeled by an effective $S=1/2$ Heisenberg chain with and without coupling.

\section{Appendix D: Finite size scaling}
\begin{figure}[htbp]
 \centering
 \includegraphics[width=0.48\textwidth]{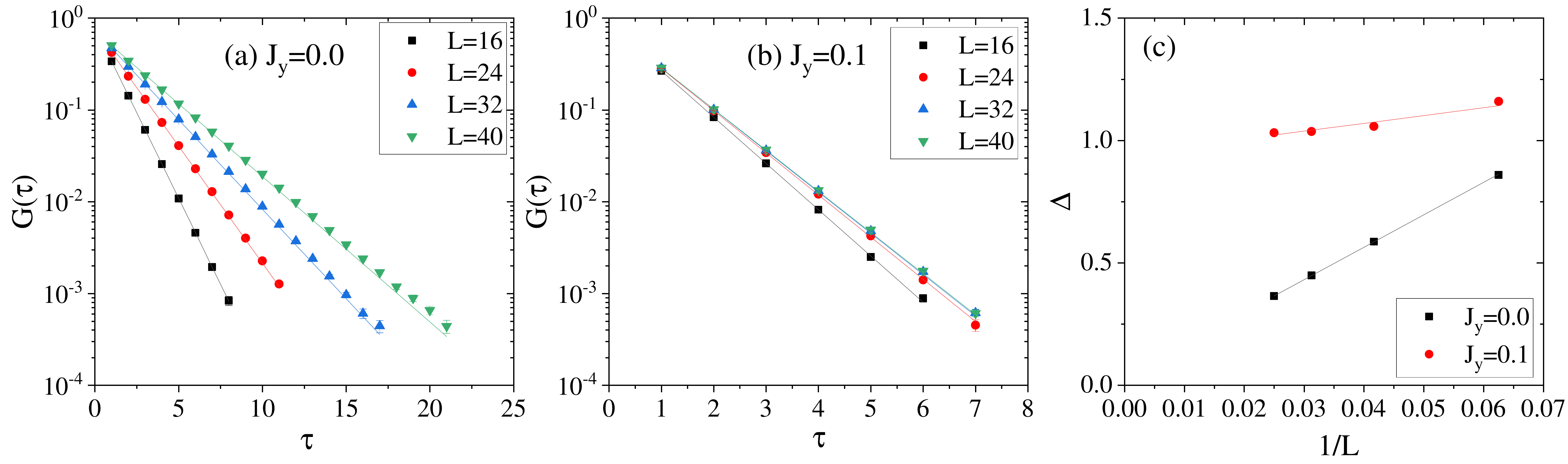}
 \caption{The finite-size scaling of the entanglement spectra gap for Fig.\ref{fig-lattice} (c) with $J_{s}=0.3$, $\beta=50$ and $\beta_{A}=2L$. (a) Imaginary-time correlation function $G(\tau)$ at $q=\pi$ with $J_{y}=0.0$. (b) Imaginary-time correlation function $G(\tau)$ at $q=\pi$ with $J_{y}=0.1$. (c)Different sizes of entanglement spectra gap with $J_{y}=0.0$ and 0.1. }
 \label{Fig.es-gap}
 \end{figure}

Although the ES gap is magnified $\beta$ times, it should satisfy the finite-size scaling of the gap. 
Imaginary-time correlations $G(\tau)$ at $q=\pi$ are measured with $J_{y}=0.0$ and 0.1 at $J_s=0.3$, and we perform the finite-size scaling analysis of the ES gap in Fig.\ref{Fig.es-gap}. In this calculation, the total system size is from 16 to 40.
These gap values are obtained from the fitting of $G(\tau)$ that should satisfy $G(\tau)=be^{-\Delta \tau}$, where $\Delta$ refers to the gap. The ES gap should obey the finite-size scaling form $\Delta=\Delta(\infty)+bL^{-1}$, where $b$ is the fitting parameters and $\Delta(\infty)$ is the ES gap in the thermodynamic limit. As Fig.\ref{Fig.es-gap} shows, it is clear that all $G(\tau)$ decay exponentially and are fitted well by the gap formula which indicates $\Delta \propto L^{-1}$. The fitting of $\Delta(\infty)$ is $0.0332(3)$ at $J_{y}=0.0$ and $0.943(1)$ at $J_{y}=0.1$. If the amplification effect is considered, the real gap $\Delta(\infty)/\beta$ at $J_{y}=0.0$ is about $6.65\times 10^{-4}$ close to zero, which agrees well with the fact that the ES is gapless. These fitting results further prove our prediction that the ES gap has been magnified $\beta$ times and explain why it looks so large at small $J_{y}$.

\section{Appendix E: Imaginary-time correlation function }
\begin{figure}[htbp]
 \centering
 \includegraphics[width=0.48\textwidth]{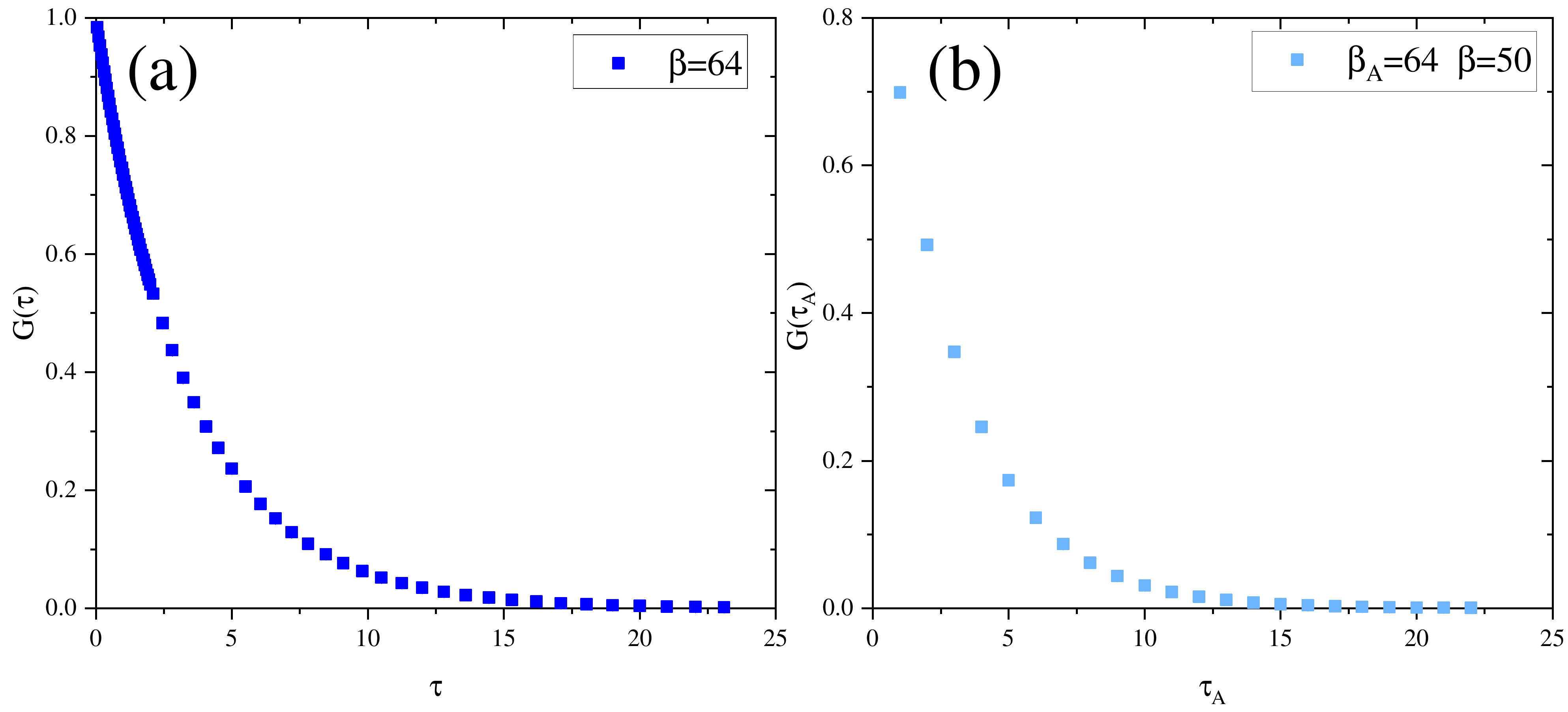}
 \caption{Imaginary-time correlation function $G(\tau)$ of the real Hamiltonian (a) and entanglement Hamiltonian (b) at $q=\pi$ with the perturbation $J_s=0.3$ on the boundary of subsystem A, as Fig.\ref{fig-lattice}(a) and (b) shown.}
 \label{Fig.es-gtau}
 \end{figure}
 The inverse temperature $\beta$ should be large enough to make the system reach the ground state like $\beta \sim L $. As Fig.\ref{Fig.es-gtau}(a) shows, imaginary-time correlation function $G(\tau)$ decays to zero with $\beta=2L=64$, which is large enough to probe the ground state properties and spin excitation.
 The number of replica $\beta_A$ should be also large enough to probe the ground state of the entanglement Hamiltonian, which is the same as the set of $\beta \sim L$. When $\beta_A=2L$, $G(\tau_A)$ also decays to zero at large $\tau$ as Fig.\ref{Fig.es-gtau}(b) shows, which guarantees that we can probe the ground state in this $\beta_A$. 

%

\end{document}